\documentclass[b5paper,twoside,reqno]{bjp}
\usepackage{cite}
\usepackage{latexsym}
\usepackage{graphicx}
\usepackage{amssymb,amsmath,amsfonts,amsbsy, amsthm, xspace, latexsym, amscd, enumitem}
\usepackage[utf8]{inputenc}
\usepackage[T1]{fontenc}
\usepackage[english]{babel}
\usepackage{times}
\usepackage{url,bm}
\usepackage{float}

\newcommand\rf[1]{(\ref{eq:#1})}
\newcommand\lab[1]{\label{eq:#1}}
\newcommand\nonu{\nonumber}
\newcommand\br{\begin{eqnarray}}
\newcommand\er{\end{eqnarray}}
\newcommand\be{\begin{equation}}
\newcommand\ee{\end{equation}}

\newcommand\lb{\lbrack}
\newcommand\rb{\rbrack}

\newcommand\bv{\bigm\vert}               

\newcommand\bc{\begin{center}}
\newcommand\ec{\end{center}}

\renewcommand\d{\delta}

\newcommand\vareps{\varepsilon}

\newcommand\h{\frac{1}{2}}
\renewcommand\k{\kappa}
\renewcommand\l{\lambda}
\renewcommand\L{\Lambda}
\newcommand\m{\mu}
\newcommand\n{\nu}
\newcommand\om{\omega}

\renewcommand\P{\Phi}
\newcommand\pa{\partial}

\newcommand\pr{\prime}

\newcommand\s{\sigma}

\renewcommand\th{\theta}

\newcommand\cC{{\mathcal C}}

\newcommand\cL{{\mathcal L}}

\newcommand{\ct}[1]{\cite{#1}}

\newcommand\PRD[3]{\textsl{Phys. Rev.} \textbf{D#1}, #3 (#2)}

\newcommand\CQG[3]{\textsl{Class. Quantum Grav.} \textbf{#1}, #3 (#2)}
\newcommand\JMP[3]{\textsl{J. Math. Phys.} \textbf{#1}, #3 (#2)}

\newcommand\PR[3]{\textsl{Phys. Reports} \textbf{#1}, #3 (#2)}

\newcommand\IJMPA[3]{\textsl{Int. J. Mod. Phys.} \textbf{A#1}, #3 (#2)}
\newcommand\IJMPD[3]{\textsl{Int. J. Mod. Phys.} \textbf{D#1}, #3 (#2)}

\newcommand\JPA[3]{\textsl{J. Physics} \textbf{A#1}, #3 (#2)}

\makeindex

\begin{document}

\sloppy \raggedbottom

\title{Four-Dimensonal Gauss-Bonnet Gravity Without Gauss-Bonnet Coupling to Matter
-- Spherically Symmetric Solutions, Domain Walls and Spacetime Singularities}

\runningheads{Four-Dimensonal Gauss-Bonnet Gravity}{E. Guendelman, 
E. Nissimov and S. Pacheva}

\begin{start}

\coauthor{Eduardo Guendelman}{1,2,3}, \coauthor{Emil Nissimov}{4},
\coauthor{Svetlana Pacheva}{4}

\address{Department of Physics, Ben-Gurion Univ. of the Negev, \\
Beer-Sheva 84105, Israel}{1}

\address{Bahamas Advanced Study Institute and Conferences, 4A Ocean Heights, \\
Hill View Circle, Stella Maris, Long Island, The Bahamas}{2}

\address{Frankfurt Institute for Advanced Studies, Giersch Science Center, \\
Campus Riedberg, Frankfurt am Main, Germany}{3}

\address{Institute of Nuclear Research and Nuclear Energy, \\
Bulgarian Academy of Sciences, Sofia 1784, Bulgaria}{4}

\begin{Abstract}
We discuss a new extended gravity model in ordinary $D=4$ spacetime
dimensions, where an additional term in the action involving Gauss-Bonnet 
topological density is included without the need to couple it to matter 
fields unlike the case of ordinary $D=4$ Gauss-Bonnet gravity models. Avoiding
the Gauss-Bonnet density becoming a total derivative is achieved by employing 
the formalism of metric-independent non-Riemannian spacetime volume-forms. 
The non-Riemannian volume element triggers {\em dynamically} the Gauss-Bonnet
scalar to become an arbitrary integration constant on-shell. We describe in some
detail the class of static spherically symmetric solutions of the above
modified $D=4$ Gauss-Bonnet gravity including solutions with deformed
(anti)-de Sitter geometries, black holes, domain walls and
Kantowski-Sachs-type universes. Some solutions exhibit physical spacetime
singular surfaces {\em not hidden} behind horizons and bordering whole forbidden 
regions of space. Singularities can be avoided by pairwise matching of two
solutions along appropriate domain walls. For a broad class of solutions 
the corresponding matter source is shown to be a special form of nonlinear 
electrodynamics whose Lagrangian $L(F^2)$ is a {\em non-analytic} function of 
$F^2$ (the square of Maxwell tensor $F_{\m\n}$), \textsl{i.e.}, $L(F^2)$ is
{\em not} of Born-Infeld type.
\end{Abstract}


\end{start}

\section{Introduction}
\label{intro}

In the last decade or so a host of problems of primary importance in 
cosmology (problems of dark energy and dark matter), quantum field theory 
in curved spacetime (renormalization in higher loops) and string theory
(low-energy effective field theories) motivated a very active development of
extended gravity theories as alternatives/generalizations of the standard
Einstein General Relativity (for detailed accounts, see 
Refs.\ct{extended-grav,extended-grav-book,odintsov-1,odintsov-2} and
references therein).

One possible approach towards alternative/extended theories to General Relativity 
is to employ the formalism of non-Riemannian spacetime volume-forms
(alternative metric-independent generally covariant volume elements or 
spacetime integration measure densities) in the pertinent Lagrangian actions,  
defined in terms of auxiliary antisymmetric tensor gauge fields of maximal rank,
instead of the canonical Riemannian volume element given by the square-root of 
the determinant of the Riemannian metric. The systematic geometrical formulation
of the non-Riemannian volume-form approach was given in 
Refs.\ct{susyssb-1,grav-bags}, which is an extension of the 
originally proposed method \ct{TMT-orig-1,TMT-orig-2}. 

This formalism is the basis for constructing a series of extended gravity-matter
models describing unified dark energy and dark matter scenario \ct{EJP},
quintessential cosmological models with gravity-assisted and inflaton-assisted
dynamical generation or suppression of electroweak spontaneous symmetry
breaking and charge confinement \ct{grf-essay,varna-17,bpu-10}, and a novel 
mechanism for the supersymmetric Brout-Englert-Higgs effect in supergravity 
\ct{susyssb-1}.


Let us recall that in standard generally-covariant theories (with actions of
the form $S=\int d^D\! x \sqrt{-g} \cL$) the standard Riemannian spacetime 
volume-form $\om$ is defined through the ``D-bein''
(frame-bundle) canonical one-forms $e^A = e^A_\m dx^\m$ ($A=0,\ldots ,D-1$),
related to the Riemannian metric ($g_{\m\n} = e^A_\m e^B_\n \eta_{AB}$ , 
$\eta_{AB} \equiv {\rm diag}(-1,1,\ldots,1)$):
\br
\om = e^0 \wedge \ldots \wedge e^{D-1} = \det\Vert e^A_\m \Vert\,
dx^{\m_1}\wedge \ldots \wedge dx^{\m_D} \; , 
\nonu
\er
so that the Riemannian volume element reads:
\be
\Omega \equiv \det\Vert e^A_\m \Vert\, d^D x = 
\sqrt{-\det\Vert g_{\m\n}\Vert}\, d^D x \; .
\lab{omega-riemannian}
\ee

Instead of $\sqrt{-g}$ we will employ below a different
alternative {\em non-Riemannian} volume element 
given by a non-singular {\em exact} $D$-form $\om = d \cC$ where:
\be
\cC = \frac{1}{(D-1)!} C_{\m_1\ldots\m_{D-1}} 
dx^{\m_1}\wedge\ldots\wedge dx^{\m_{D-1}} \; ,
\lab{C-form}
\ee
so that the {\em non-Riemannian} volume element becomes:
\be
\Omega \equiv \Phi(C) d^D x = 
\frac{1}{(D-1)!}\vareps^{\m_1\ldots\m_D}\, \pa_{\m_1} C_{\m_2\ldots\m_D} d^D x\; .
\lab{Phi-D}
\ee
Here $C_{\m_1\ldots\m_{D-1}}$ is an auxiliary rank $(D-1)$ antisymmetric tensor 
gauge field. $\Phi(C)$ is in fact the density of the dual of the rank 
$D$ field-strength
$F_{\m_1 \ldots \m_D} = \frac{1}{(D-1)!} \pa_{\lb\m_1} C_{\m_2\ldots\m_D\rb}
= - \vareps_{\m_1 \ldots \m_D} \P (C)$. Like $\sqrt{-g}$, $\P (C)$
similarly transforms as scalar density under general coordinate reparametrizations.

Now, we observe that if we replace the usual Riemannian volume element density
$\sqrt{-g}$ with a non-Riemannian one $\P(C)$ \rf{Phi-D} in the Lagrangian 
action integral over the 4-dimensional Gauss-Bonnet scalar 
$\int d^4 x\,\P(C)\,R^2_{\rm GB}$ 
(cf. Eqs.\rf{TMT-GB}-\rf{GB-def} below), then the latter
will cease to be a total derivative in $D=4$. In this way we will avoid the
necessity to couple $R^2_{\rm GB}$ in $D=4$ directly to matter fields 
or to use nonlinear functions of $R^2_{\rm GB}$
unlike the usual $D=4$ Einstein-Gauss-Bonnet gravity. 
For reviews of the latter, see Refs.\ct{GB-grav-3,GB-grav-4}; 
for recent discussions of
Gauss-Bonnet cosmology, see
Refs.\ct{GB-grav-cosmolog-1}-\ct{GB-grav-cosmolog-10},
and references therein.

Our non-standard $D=4$ Gauss-Bonnet gravity with a Gauss-Bonnet action term 
$\int d^4 x\,\P(C) R^2_{\rm GB}$ has the following principal properties:

\begin{itemize}
\item
The equation of motion w.r.t. auxiliary tensor gauge field 
$C_{\m_1\ldots\m_{D-1}}$ defining $\P(C)$ \rf{Phi-D} dynamically triggers the 
Gauss-Bonnet scalar $R^2_{\rm GB}$ to be on-shell an arbitrary integration constant
(Eq.\rf{GB-const} below).
\item
Now the composite field $\chi = \frac{\P(C)}{\sqrt{-g}}$ appears as an 
additional physical field degree of freedom related to the geometry of spacetime.
Let us note that the latter is in sharp
contrast w.r.t. other extended gravity-matter models constructed in terms of 
(one or several) non-Riemannian volume-forms 
\ct{susyssb-1,grav-bags,EJP,grf-essay,varna-17,bpu-10}, where we start within 
the first-order (Palatini) formalism and where
composite fields of the type of $\chi$ 
(ratios of non-Riemannian to Riemannian volume element densities)
turn out to be (almost) pure gauge (non-propagating) degrees of freedom, the
only remnants being the appearance of some further free integration constants.
\end{itemize}

The dynamically triggered constancy of $R^2_{\rm GB}$ in our modified
$D=4$ Gauss-Bonnet gravity has several interesting implications for cosmology
\ct{4D-GB-cosmolog}, in particular, the additional degree of freedom $\chi$ 
absorbing completely the effect of the matter dynamics within the 
Friedmann-Lemaitre-Robertson-Walker formalism. The above properties are the
most significant differences of the present approach w.r.t. the approach in
several recent papers \ct{myrzakulov,vagenas,radinschi}, which extensively 
study static spherically symmetric solutions in gravitational theories in 
the presence of a constant Gauss-Bonnet scalar. In the latter papers the 
constancy of the Gauss-Bonnet scalar is {\em imposed as an additional condition 
on-shell} beyond the standard equations of motion resulting from an action
principle. Therefore, the full set of equations (equations of motion plus
the {\em ad hoc} imposed constancy of $R^2_{\rm GB}$) in the latter papers
is {\em not equivalent} to the full set of equations of motion in the
present modified $D=4$ Gauss-Bonnet gravity based on the non-Riemannian
spacetime volume-form formalism.

The plan of the paper is as follows. After presenting in Section 2 the basics of 
the non-Riemannian volume-form formulation of modified $D=4$ Gauss-Bonnet gravity,
in Section 3 we describe the general properties of the whole class of static
spherically symmetric solutions for the various values of the pertinent free 
integration constants. 

In Section 4 we analyze in some detail the domains of
definition of the static spherically symmetric metrics and the locations of 
physical spacetime singularities, domain walls and horizons.
The spacetime singularities of the modified $D=4$ Gauss-Bonnet gravity
are constant $r=r_{*}$ surfaces bordering whole forbidden space regions of finite
or infinitely large extent where the metric becomes complex. 
Most of these spacetime singularities are {\em not hidden} behind horizons.
They resemble the so called branch singularities at finite $r$ of static
spherically symmetric solutions in higher-dimensional ($D\geq 5$)
Einstein-Maxwell-Gauss-Bonnet gravity \ct{torii-maeda} where the 
higher-dimensional quadratic curvature invariants exhibit the same singular 
behaviour near $r_{*}$ as in the present case (cf. Eq.\rf{R-diverge-2} below).

Section 5 contains the graphical representations of the whole class of static
spherically symmetric solutions. In Section 6 we briefly illustrate how to
avoid spacetime singularities via pairwise matching of two solutions along
appropriate domain wall. In the last discussion Section 7 we add
some comments and conclusions.

\section{Gauss-Bonnet Gravity in $D=4$ With a Non-Riemannian Volume Element}

We propose the following self-consistent action of $D=4$ Gauss-Bonnet
gravity without the need to couple the Gauss-Bonnet scalar to some matter
fields
(for simplicity we are using units with the Newton constant $G_N = 1/16\pi$):
\be
S = \int d^4 x \sqrt{-g} \Bigl\lb R + L_{\rm matter}\Bigr\rb
+ \int d^4x \,\P (C)\, R_{\rm GB}^2 \; .
\lab{TMT-GB}
\ee
Here the notations used are as follows:
\begin{itemize}
\item
$R_{\rm GB}^2$ denotes the Gauss-Bonnet scalar:
\be
R_{\rm GB}^2 \equiv R^2 - 4 R_{\m\n} R^{\m\n} + R_{\m\n\k\l} R^{\m\n\k\l} \; .
\lab{GB-def}
\ee
\item
$\P (C)$ denotes a non-Riemannian volume element density defined as a scalar density
of the dual field-strength of an auxiliary antisymmetric tensor gauge field of 
maximal rank $C_{\m\n\l}$:
\be
\P (C) = \frac{1}{3!} \vareps^{\m\n\k\l} \pa_\m C_{\n\k\l} \; .
\lab{PC-def}
\ee
Let us particularly stress that, although we stay in $D=4$ spacetime
dimensions and although we {\em don't couple} the Gauss-Bonnet scalar 
\rf{GB-def} to the matter fields, the last term in \rf{TMT-GB}
thanks to the presence of the non-Riemannian volume element \rf{PC-def} is 
non-trivial ({\em not} a total derivative as with the ordinary Riemannian
volume element $\sqrt{-g}$)) and yields a non-rivial contribution to the
Einstein equations (Eqs.\rf{einstein-eqs} below).
\item
As we will see in what follows, the specific form of the matter Lagrangian 
$L_{\rm matter}$ in \rf{TMT-GB} will depend on the specific class of 
static spherically symmetric (SSS) solutions we are looking for 

$\phantom{aa}$(i) For a broad class of SSS solutions specified below 
(see Eqs.\rf{L-F-A-1}-\rf{L-F-A-3} below)
$L_{\rm matter}$ will be required by consistency of the equations of motion
to be a Lagrangian of a {\em nonlinear electrodynamics} $L(F^2)$:
\be
L_{\rm matter} = L(F^2) \quad ,\quad
F^2 \equiv F_{\m\n} F_{\k\l} g^{\m\k} g^{\n\l} \quad ,\;\;
F_{\m\n} = \pa_\m A_\n - \pa_\n A_\m \; .
\lab{NL-ED}
\ee
An important property of $L(F^2)$ we prove below is that the latter must be a 
{\em non-analytic} function of $F^2$.

$\phantom{aa}$(ii) A special narrow class of SSS solutions (Eq.\rf{hedgehog-SSS}
below) will require ``hedgehog'' scalar field matter source with a Lagrangian of a 
$O(3)$ nonlinear sigma-model ($\l$ being a Lagrange multiplier)
with a ``hedgehog'' solution \ct{eduardo-rabinowitz}:
\be
L_{\rm matter} = -\h g^{\m\n}\pa_\m \vec{\phi}. \pa_\n \vec{\phi}
-\l \bigl(\vec{\phi}.\vec{\phi} - v^2\bigr) \quad ,\quad
\vec{\phi} = \pm v \hat{r} \; .
\lab{NL-sigma}
\ee
with $\hat{r}$ -- unit radial vector.

$\phantom{aa}$(iii) Another set of SSS solutions require matter sources lacking 
explicit Lagrangian action formulation, including additional thin-shell (brane) ones
describing domain walls (Eqs.\rf{A-DW}-\rf{T-brane} below). 
\end{itemize}

We now have three types of equations of motion resulting from the action
\rf{TMT-GB}: 
\begin{itemize}
\item
Einstein equations w.r.t. $g^{\m\n}$ where we employ the definition for a
composite field:
\be
\chi \equiv \frac{\P(C)}{\sqrt{-g}} 
\lab{chi-def}
\ee
representing the ratio of the non-Riemannian to the standard Riemannian 
volume element densities:
\br
R_{\m\n} - \h g_{\m\n}R = \h T_{\m\n} - \h g_{\m\n} \chi R_{\rm GB}^2
+ 2R \nabla_\m \nabla_\n \chi 
\nonu \\
+ 4 \Box \chi \bigl( R_{\m\n} - \h g_{\m\n}R \bigr)
- 4 R_{\m}^{\rho} \nabla_\rho \nabla_\n \chi
- 4 R_{\n}^{\rho} \nabla_\rho \nabla_\m \chi 
\nonu \\
+ 4 g_{\m\n} R^{\rho \s} \nabla_\rho \nabla_\s \chi 
- 4 g^{\k\rho} g^{\l\s} R_{\m\k\n\l} \nabla_\rho \nabla_\s \chi \; , \;\;\; 
\lab{einstein-eqs}
\er
where 
$T_{\m\n} = g_{\m\n} L_{\rm matter} - 2\frac{\pa}{\pa g^{\m\n}}L_{\rm matter}$ 
is the relevant standard matter energy-momentum tensor.
In particular, for the nonlinear electrodynamics:
\be
T_{\m\n} = g_{\m\n} L(F^2) - 4 L^{\pr}(F^2) F_{\m\k} F_{\n\l} g^{\k\l} \; ,
\lab{EM-tensor}
\ee
where $L^{\pr}(F^2) \equiv \frac{\pa L}{\pa F^2}$, and for the scalar
``hedgehog'' field ${\vec\phi}$:
\be
T_{\m\n} = \pa_\m \vec{\phi}.\pa_\n \vec{\phi} 
- \h g_{\m\n} g^{\k\l}\pa_\k \vec{\phi}. \pa_\l \vec{\phi} \; .
\lab{hedgehog-tensor}
\ee
\item
The equations of motion w.r.t. scalar ``hedgehog'' field ${\vec\phi}$ and the
nonlinear gauge field  have the standard form (they are not affected by the 
presence of the Gauss-Bonnet term):  
\br
\Box {\vec\phi} -\l {\vec\phi} = 0 \quad ,\quad \vec{\phi}.\vec{\phi} - v^2 = 0 \; ,
\lab{vp-eq} \\
\nabla_\n \bigl( L^{\pr}(F^2) F^{\m\n}\bigr) = 0  \; .
\lab{NL-ED-eq}
\er
\item
The crucial new feature are the equations of motion w.r.t. auxiliary
non-Riemannian volume element tensor gauge field $C_{\m\n\l}$:
\be
\pa_\m R_{\rm GB}^2 = 0 \quad \longrightarrow \quad
R_{\rm GB}^2 = 24 M = {\rm const} \; ,
\lab{GB-const}
\ee
where $M$ is an {\em arbitrary dimensionful integration constant} and 
the numerical factor 24 in \rf{GB-const} is chosen for later convenience.
\end{itemize}

The dynamically triggered constancy of the Gauss-Bonnet scalar \rf{GB-const}
comes at a price as we see from the generalized Einstein
Eqs.\rf{einstein-eqs} -- namely, now the composite field 
$\chi = \frac{\P(C)}{\sqrt{-g}}$ appears as an {\em additional} physical field
degree of freedom.

In what follows we will see that when considering SSS solutions we can 
consistently ``freeze'' the composite field $\chi = {\rm const}$ so that 
all terms on the r.h.s. of \rf{einstein-eqs} with derivatives of the composite 
field $\chi$ will vanish.
Thus, we are left with an {\em overdetermined} system of equations:
\be
R_{\m\n} - \h g_{\m\n}R = \h T_{\m\n} - g_{\m\n} 12 \chi M \quad ,\quad
R_{\rm GB}^2 = 24 M \; ,
\lab{einstein-eqs-chi-const}
\ee
plus the matter field equations of motion \rf{vp-eq}-\rf{NL-ED-eq}
determining $T_{\m\n}$.


\section{Static Spherically Symmetric Solutions with a Dynamically Constant 
Gauss-Bonnet Scalar -- General Properties}
\label{SSS}

Let us now consider the system \rf{einstein-eqs-chi-const} with a
static spherically symmetric (SSS) ansatz for the metric:
\be
ds^2 = - A(r) dt^2 +\frac{dr^2}{A(r)} +
r^2 \bigl(d\th^2 +\sin^2\th d\phi^2\bigr) \; .
\lab{spherical-symm}
\ee
Inserting \rf{spherical-symm} in \rf{GB-const} we have:
\be
R_{\rm GB}^2 = 24M \quad\longrightarrow \quad 
\frac{2}{r^2} \frac{d^2}{dr^2}\Bigl\lb \bigl(A(r)-1\bigr)^2\Bigr\rb = 24M \;,
\lab{GB-const-spherical}
\ee
which yields the following general solution for $A(r)$ already noted in
Ref.\ct{vagenas}:
\be
A(r) = 1 \pm \sqrt{P_4(r)} \quad ,\quad P_4(r) \equiv Mr^4 + c_1 r + c_0
\lab{A-sol}
\ee
with $c_{0,1}$ together with $M$ representing three {\em a priori} arbitrary
integration constants. Let us stress that the
gravity solution \rf{A-sol} is not affected by the matter sources.

For the SSS ansatz \rf{spherical-symm} the Ricci tensor components and the
scalar curvature read:
\br
R^0_0 = R^r_r= - \frac{1}{2r^2} \pa_r\bigl( r^2 \pa_r A\bigr)  \quad ,\quad 
R^i_j= - \d^i_j \Bigl\lb \frac{1}{r^2} (A-1) + \frac{1}{r} \pa_r A\Bigr\rb \; ,
\lab{Ricci-SSS} \\
R = -2\Bigl\lb \frac{1}{r^2} (A-1) + \frac{1}{r} \pa_r A\Bigr\rb 
- \frac{1}{r^2} \pa_r\bigl( r^2 \pa_r A\bigr) \; ,
\lab{R-SSS}
\er
whereupon the Einstein equations in \rf{einstein-eqs-chi-const} become:
\br 
\frac{1}{r^2} (A-1) + \frac{1}{r} \pa_r A = \h T^0_0 - 12\chi M \; ,
\lab{Einstein-SSS-1} \\
\frac{1}{2r^2} \pa_r\bigl( r^2 \pa_r A\bigr) = \h T^\th_\th - 12\chi M \; .
\lab{Einstein-SSS-2}
\er
Consistency of SSS Einstein equations \rf{Einstein-SSS-1}-\rf{Einstein-SSS-2}
requires for the components of the matter energy momentum tensor:
\be
T^0_0 = T^r_r \quad ,\quad T^\th_\th = T^\phi_\phi \quad, \quad 
{\mathrm rest} = 0 \; .
\lab{EM-tensor-cond}
\ee

Conditions \rf{EM-tensor-cond} are fulfilled for the SSS solutions in 
nonlinear electrodynamics ($F_{0r}= F_{0r} (r)$ being the only surviving 
component of $F_{\m\n}$; $F^2 = -2 F^2_{0r}$):
\be
T^0_0 = T^r_r = L(F^2) + 4 F_{0r}^2 L^\pr (F^2) \quad ,\quad
T^\th_\th = T^\phi_\phi = L(F^2) \; ,
\lab{EM-tensor-SSS-NLED}
\ee
where Eqs.\rf{NL-ED-eq} reduce to:
\be
\pa_r \bigl(r^2 F_{0r} L^\pr (F^2)\bigr) = 0 \;\;\; \longrightarrow \;\;\;
F_{0r} L^\pr (F^2) = \frac{q}{16\pi\,r^2} \; ,
\lab{NL-ED-eq-SSS}
\ee
$q$ indicating the electric charge.

Conditions \rf{EM-tensor-cond} are fulfilled as well as for the SSS ``hedgehog''
solution of \rf{NL-sigma}:
\be
T^0_0 = T^r_r = - v^2/r^2 \quad ,\quad T^\th_\th = T^\phi_\phi = 0 \; .
\lab{EM-tensor-SSS-hedgehog}
\ee

In the case of nonlinear electrodynamics source, combining
\rf{Einstein-SSS-1}-\rf{Einstein-SSS-2} with 
\rf{EM-tensor-SSS-NLED}-\rf{NL-ED-eq-SSS} we obtain an exact expression
for the radial electric field ($E_r \equiv - F_{0r}$) in terms of the 
metric function \rf{A-sol}:
\be
F_{0r} = - \frac{4\pi\,r^2}{q} \Bigl\lb \pa_r^2 A - \frac{2}{r^2}(A-1)\Bigr\rb \; .
\lab{F-sol}
\ee

Let us note the following two obvious well-defined non-trivial solutions for 
$A(r)$ \rf{A-sol} satisfying \rf{Einstein-SSS-1}-\rf{Einstein-SSS-2}:

\begin{itemize}
\item
For $\bigl( M>0\,,\, c_0=c_1=0\bigr)$ \rf{A-sol} becomes the standard (anti)-de
Sitter solution $A(r) = 1 \pm \sqrt{M}r^2$, where $T^0_0 = T^\th_\th = 0$ and 
$\chi = \mp \frac{1}{4\sqrt{M}}$.
\item
$A(r)$ \rf{A-sol} becomes for:
\be
\bigl( M=0\,,\, c_1=0 \,,\, c_0 = v^4/4\bigr) \;\; \longrightarrow \;\;
A(r) = 1 - v^2/2 = {\rm const} \; ,
\lab{hedgehog-SSS}
\ee
the (minus) $00$-component of the metric generated by the SSS energy-momentum tensor
\rf{EM-tensor-SSS-hedgehog} of the ``hedgehog'' scalar field \rf{NL-sigma}.
\end{itemize}

Before proceeding let us stress that:
\begin{itemize}
\item
The solutions $A(r)$ \rf{A-sol} will have well-defined large $r$ asymptotics 
only for $M>0$ (see \rf{A-sol-asymptot} below), or for $(M=0\,,\,c_1>0)$.
In the case $M<0$ or $(M=0\,,\,c_1<0)$ the large $r$ region will be inaccessible 
(forbidden region) since $A(r)$ becomes complex there,
\textsl{i.e.} spacetime does not exist there.
\item
Similarly, for all solutions $A(r)$ \rf{A-sol} with $c_0<0$ the region of
small $r$, where $A(r)$ becomes complex, will be inaccessible (forbidden region).
\end{itemize}

Asymptotically, for large $r$ (cf. \ct{vagenas}) $A(r)$ \rf{A-sol} with $M>0$
becomes:
\be
A(r) \simeq 1 \pm \Bigl( \sqrt{M} r^2 + \frac{c_1}{2\sqrt{M}\,r} 
+ \frac{c_0}{2\sqrt{M}\,r^2}\Bigr) + {\rm O}(r^{-4}) \; ,
\lab{A-sol-asymptot}
\ee
so that asymptotially \rf{A-sol-asymptot} can be viewed as 
Reissner-Nordstr{\"o}m-(anti)-de Sitter metric 
$A_{\rm RN-(A)dS}(r)=1 \pm \frac{\L}{3} r^2 - \frac{2m}{r} +
\frac{q^2}{(8\pi)^2 r^2}$ upon the following
identification of the signs and values of the free integration constants
$(M,c_1,c_0)$:
\be
\sqrt{M} = \L/3 \;\; ,\;\; c_1 = \mp 4m\sqrt{M} \;\; ,
\;\; c_0 = \pm \frac{q^2}{32\pi^2}\sqrt{M} \; ,
\lab{param-ident}
\ee
where the upper/lower signs in \rf{param-ident} and below refer to 
anti-de Sitter/de Sitter asymptotics. 

Now, let us insert \rf{A-sol-asymptot} in \rf{Einstein-SSS-2} with a nonlinear
electrodynamics source:
\be
L(F^2) - 24\chi M = \pa_r^2 A + \frac{2}{r} \pa_r A \simeq
\pm \Bigl(6\sqrt{M} + \frac{c_0}{\sqrt{M}\,r^4} - \frac{3 c_1^2}{2 M^{3/2}r^6} +
{\rm O}(r^{-7})\Bigr) \; ,
\lab{L-asymptot}
\ee
and compare with the large $r$ asymptotics of $F_{0r}$ upon inserting
\rf{A-sol-asymptot} in \rf{F-sol}:
\be
F_{0r} \simeq \mp \frac{4\pi}{q} \Bigl(\frac{2c_0}{\sqrt{M}\,r^2}  
- \frac{9c_1^2}{4M^{3/2}r^4} + {\rm O}(r^{-5})\Bigr)
\lab{F-asymptot}
\ee
(noting that for SSS configurations $F^2 = -2 F^2_{0r}$). Thus,  we obtain
again $\chi = \mp \frac{1}{4\sqrt{M}}$ as for the pure (anti)-de Sitter
solution, but more importantly, we explicitly find that for weak 
electromagnetic fields the nonlinear electrodynamics Lagrangian is a 
{\em non-analytic} function of $F^2$:
\br
L(F^2)=-\frac{1}{4}F^2 + c_2\sqrt{-F^2}\,(-F^2) + {\rm O}\bigl((-F^2)^2\bigr)
\lab{L-nonanalytic}\\
c_2 \equiv \pm \frac{3\sqrt{2}\,m^2}{\sqrt{M}} \Bigl(\frac{4\pi}{q}\Bigr)^3
\lab{c2}
\er
using the parameter identification \rf{param-ident}, in other words $L(F^2)$
{\em is not of Born-Infeld type}.

Let us consider again the system of three equations \rf{A-sol}, \rf{L-asymptot} 
and \rf{F-sol}:
\br
A(r) = 1 \pm \sqrt{Mr^4 + c_0 + c_1 r} \; ,
\lab{L-F-A-1} \\
F_{0r}=-\frac{4\pi\,r^2}{q} \Bigl\lb \pa_r^2 A -\frac{2}{r^2}(A-1)\Bigr\rb \; ,
\lab{L-F-A-2}\\
L(F^2) = 24\chi M + \pa_r^2 A + \frac{2}{r} \pa_r A 
\nonu \\
= 24\chi M  - \frac{q}{4\pi\,r^2}F_{0r} + 2\Bigl\lb \frac{1}{r}\pa_r A 
+ \frac{1}{r^2}(A-1)\Bigr\rb \; .
\lab{L-F-A-3}
\er
In principle \rf{L-F-A-1}-\rf{L-F-A-3} allows to determine the full
nonlinear, and non-analytic as we proved in \rf{L-nonanalytic}, 
functional expression of $L(F^2)$ by first expressing $r$ as implicit function 
of $F_{0r}$ from \rf{L-F-A-1}-\rf{L-F-A-2} and then substituting in \rf{L-F-A-3}
using \rf{L-F-A-1} (recall $F^2 = -2 F_{0r}^2$).

Now, an important remark regarding the matter sources in \rf{einstein-eqs-chi-const}
is in order.

\begin{itemize}
\item
According to \rf{param-ident} and \rf{L-asymptot} only metrics
\rf{spherical-symm} with $A(r)$ \rf{A-sol}, whose parameters are of the form 
$(M>0,c_0>0)$ for anti-de Sitter asymptotics (upper sign in \rf{A-sol})
or $(M>0,c_0<0)$ for de Sitter-like asymptotics (lower sign in
\rf{A-sol}), will have nonlinear electrodynamical matter source.
\item
In what follows we will concentrate on solutions for the SSS metric $A(r)$
\rf{A-sol} with de Sitter-like asymptotics. Thus, while being generated by 
nonlinear electrodynamics source (for $(M>0,c_0<0)$) the pertinent $A(r)$ 
\rf{A-sol} will become complex for small $r$ as already pointed out above.
Therefore, for all SSS solutions \rf{A-sol} with de Sitter-like large $r$ 
asymptotics, whose matter source is nonlinear electrodynamics,
the region of small $r$ will be a forbidden one. In particular,
in this case (for $(M>0,c_0<0)$) there are {\em no} black hole solutions.
For $(M>0,c_0<0)$ there is only one SSS solution with a horizon -- 
with de Sitter-like geometry outside the forbidden small $r$ region 
(see Fig.10 below).
The above results conform to the non-existence theorems of Ref.\ct{bronnikov}
(see also \ct{dymnikova}) stating that for nonlinear electrodynamics source with 
$L(F^2) \sim F^2$ at weak fields (as in \rf{L-nonanalytic}) the SSS electrically 
charged solutions cannot have a regular center at $r=0$.
\item
On the other hand, when $(M>0,c_0>0)$ the matter source for $A(r)$ \rf{A-sol} 
with de Sitter-like asymptotics will be formally again nonlinear electrodynamics
but with a purely imaginary electric charge $q$ -- recall from 
\rf{param-ident} $c_0 = - \frac{q^2}{32\pi^2}\sqrt{M}$ and compare with the
large $r$ asymptotics \rf{L-asymptot}-\rf{F-asymptot} with the lower signs. 
This is similar to the formal electromagnetic source producing the
Riessner-Nordstr{\"o}m-like metric with a negative charge-squared ($q^2 < 0$)
in Einstein-Rosen's classic 1935 paper \ct{ER-1935}.
In this latter case $(M>0,c_0>0)$ there exist black hole solutions with de
Sitter large $r$ asymptotics for $A(r)$ \rf{A-sol} -- see Figs.6,7,9 below.
\end{itemize}

\section{Domains of Definition and Horizons for the Metric Function 
$\mathbf{A(r)}$}
\label{domains-horizons}

The defining domain of $A(r)$ \rf{A-sol}, \textsl{i.e.}, for those $r$
for which $A(r)$ is real-valued, is given by the condition on the 
4-th order polynomial $P_4(r)$ under the square root in \rf{A-sol}:
\be
P_4(r) \equiv Mr^4 + c_1 r + c_0 \geq 0 \; .
\lab{defining-domain}
\ee
The intervals of $r$ where $P_4 (r) < 0$ are forbidden regions (spacetime
does not exist there since $A(r)= 1 \pm \sqrt{P_4 (r)}$ becomes complex-valued).

\begin{itemize}
\item
Simple positive roots $r_{*}>0$ of $P_4(r)$ 
(where $P_4(r) \simeq (r-r_{*})P^\pr_4 (r_{*}) + \\
{\rm O}\bigl((r-r_{*})^2 \bigr)$) :
\be
P_4(r_{*}) = 0 \;\; \longrightarrow \;\; 
\pa_r A = \pm \frac{P^\pr_4 (r)}{2\sqrt{P_4 (r)}}
\simeq \pm \frac{\sqrt{|P^\pr_4 (r_{*})|}}{2 |r-r_{*}|^{1/2}} 
\to \pm \infty 
\lab{simple-root}
\ee
for $r\to r_{*}$, 
signify the existence of a physical spacetime singularity -- \textsl{e.g.}, 
the scalar curvature $R$ \rf{R-SSS} and the quadratic curvature invariants
diverge there:
\br
R \sim \mp \frac{\sqrt{|P^\pr_4 (r_{*})}}{4|r-r_{*}|^{3/2}}
\to \mp \infty \; , 
\lab{R-diverge-1} \\
{\rm quadratic ~curvature ~invariants} = {\rm O}\bigl(|r-r_{*}|^{-3}\bigr) \; .
\lab{R-diverge-2}
\er
Similarly, also the electric field \rf{L-F-A-2} has the same singularity at 
$r=r_{*}$ as in \rf{R-diverge-1}.

$\phantom{aaa}$(i) When $P^\pr_4 (r_{*}) > 0$ the forbidden region (for de Sitter
asymptotics -- lower sign in \rf{simple-root}) is a finite-extent
internal one $(0 < r < r_{*})$ -- see Fig.10 below where 
$(M>0, c_1 ~{\rm any}, c_0<0)$ and Fig.21 below where $(M=0, c_1>0, c_0<0)$.

$\phantom{aaa}$(ii) When $P^\pr_4 (r_{*}) < 0$ the forbidden region (for de Sitter
asymptotics) is an infinite-extent external one $(r_{*} < r < \infty)$ --
see Figs.11,12,13,14 below where $(M<0, c_1 ~{\rm any}, c_0>0)$, 
and Fig.19 below where $(M=0, c_1<0, c_0>1)$.

\item
Two simple positive roots $r_{1*}>0$ and $r_{2*}>0$ of $P_4(r)$,
$r_{1*} < r_{2*}$,
with two physical spacetime singularities there (cf. 
\rf{simple-root}-\rf{R-diverge-2}).

$\phantom{aaa}$(i) For $P^\pr_4 (r_{1*}) < 0$ and $P^\pr_4 (r_{2*}) > 0$
the forbidden region (for de Sitter asymptotics) is a finite-extent intermediate  
one $(r_{1*} < r < r_{2*})$ -- see Figs.8,9 below where 
$(M>0, c_1 < -4M\bigl(c_0/3M\bigr)^{3/4},c_0>0)$.

$\phantom{aaa}$(ii) For $P^\pr_4 (r_{1*}) > 0$ and $P^\pr_4 (r_{2*}) < 0$
there are two forbidden regions (for de Sitter asymptotics):
a finite-extent internal $(0 < r < r_{1*})$ and an infinite-extent external 
$(r_{2*} < r < \infty)$. See Figs.15,16,17 below where 
$(M<0, c_1 ~{\rm any}, c_0<0)$.
\item
Double positive root $r_{\rm DW}\equiv \bigl(c_0/3M\bigr)^{1/4}$ of $P_4(r)$:
\be
P_4(r) = (r-r_{\rm DW})^2 M \bigl\lb 6r^2_{\rm DW} + 4r_{\rm DW}(r-r_{\rm DW})
+ (r-r_{\rm DW})^2\bigr\rb
\lab{P4-DW}
\ee
yield  spacetime geometry \rf{spherical-symm} with:
\be
A(r) = 1 \pm |r-r_{\rm DW}| \sqrt{M \bigl\lb 6r^2_{\rm DW} + 
4r_{\rm DW}(r-r_{\rm DW}) + (r-r_{\rm DW})^2\bigr\rb}
\lab{A-DW}
\ee
which contains a {\em domain wall} located at 
$r=r_{\rm DW}\equiv \bigl(c_0/3M\bigr)^{1/4}$ 
(see Fig.2 and Fig.7 below where 
$(M>0, c_1 = -4M\bigl(c_0/3M\bigr)^{3/4}, c_0<0)$) since while
$A(r)$ is continuous there, its derivative $\pa_r A$ has a discontinuity.
Therefore, the second derivative gets a delta-function contribution plus an
additional discontinuity at $r=r_{\rm DW}\equiv \bigl(c_0/3M\bigr)^{1/4}$:
\be
\pa_r^2 A = - \sqrt{24M} r_{\rm DW} \d (r-r_{\rm DW}) 
- \sqrt{\frac{8M}{3}} {\rm sign}(r-r_{\rm DW}) + {\rm regular} \; .
\lab{A-delta}
\ee
Eq.\rf{A-delta} through the SSS Einstein equation \rf{Einstein-SSS-2} and
Eq.\rf{F-sol} indicates the presence of a surface stress-energy tensor $S^\m_\n$ 
of an additional static charged thin-shell (brane) matter source located at 
$r=r_{\rm DW}\equiv \bigl(c_0/3M\bigr)^{1/4}$ so that \rf{Einstein-SSS-2} is
modified as:
\br
L(F^2)=24\chi M+\pa_r^2 A +\frac{2}{r}\pa_r A - {T^\th_\th}\bv_{\rm brane} \;, 
\lab{Einstein-SSS-2+brane}\\
{T^\m_\n}\bv_{\rm brane} = S^\m_\n \d (r-r_{\rm DW}) \quad ,\quad
S^0_0=S^r_r=0 \;\;,\;\; S^\th_\th = S^\phi_\phi \; ,
\lab{T-brane}
\er
whereas \rf{NL-ED-eq-SSS} is modified as:
\be
F_{0r} L^\pr (F^2) = \frac{q}{16\pi\,r^2} + 
\frac{1}{8} j^0_{\rm brane} {\rm sign}(r-r_{\rm DW}) \; ,
\lab{NL-ED-eq-SSS+brane}
\ee
with $j^0_{\rm brane}$ being the surface brane charge density.
Relations \rf{T-brane} comply with the general formalism for thin-shell
domain walls developed in Ref.\ct{BGG-87}.

Similarly, there appear the same brane stress-energy and surface charge 
contributions in the modification of \rf{F-sol}:
\br
F_{0r}\Bigl\lb \frac{q}{16\pi\,r^2} + 
\h j^0_{\rm brane}{\rm sign}(r-r_{\rm DW})\Bigr\rb 
\nonu \\
= - \frac{4\pi\,r^2}{q} \Bigl\lb \pa_r^2 A - \frac{2}{r^2}(A-1)
- {T^\th_\th}\bv_{\rm brane} \Bigr\rb \; .
\lab{F-sol+brane}
\er
Choosing the value of the brane pressure: 
\be
S^\th_\th = - \sqrt{24M} r_{\rm DW} \; ,
\lab{brane-pressure}
\ee
we exactly cancell the delta-function part in $L(F^2)$ \rf{Einstein-SSS-2+brane}
and $F_{0r}$ \rf{F-sol+brane} due to the delta-function singularity in
$\pa_r^2 A$ \rf{A-delta}, whereas the discontinuous term on the l.h.s. of
\rf{F-sol+brane} is matched by the discontinuity in $\pa_r^2 A$ \rf{A-delta}.
Let us note that Eq.\rf{brane-pressure} together with \rf{T-brane} implies
that the thin-shell matter forming the domain wall is an {\em exotic matter}
(violating null energy condition).
\item
Another class of SSS solutions for $A(r)$ with de Sitter-like asymptotics is when
$P_4(r) >1$ for all $r$, \textsl{i.e.}, $A(r) = 1 - \sqrt{P_4 (r)} <0$ for all
$r$, which means that $r$ becomes timelike whereas $t\equiv z$ becomes 
``radial-like'' 
spacelike. In this case the SSS metric \rf{spherical-symm} acquires the
following form upon introducing a new ``cosmological'' time coordinate $\xi$ 
instead of the timelike $r$:
\be
ds^2 = - d\xi^2 + |A\bigl(r(\xi)\bigr)| dz^2 + 
r^2(\xi)\bigl(d\th^2 +\sin^2\th d\phi^2\bigr) \;\; ,\;\;
\frac{dr}{d\xi} = \sqrt{|A\bigl(r(\xi)\bigr)|} \; ,
\lab{K-S}
\ee
This describes the geometry of a particular type of Kantowski-Sachs universe
\ct{K-S} -- contracting, expanding or bouncing -- depending of the values of the
free integration constants. See Figs.3,4,5 below where
$(M>0, c_1\geq -4M\bigl(c_0/3M\bigr)^{3/4}, c_0>1)$ and Fig.18 below where
$(M=0, c_1>0, c_0>1)$.
\end{itemize}

On the other hand, for de Sitter-like asymptotics (lower sign in \rf{A-sol})
there might exist one or two horizons $r_0$ of $A(r)$ provided there are
(one or two) positive roots $r_0$ of the related polynomial $Q_4 (r)$:
\br
Q_4 (r) \equiv P_4 (r) - 1 = Mr^4 + c_1 r + c_0 -1 \; ,
\lab{Q4-def} \\
Q_4 (r_0) = 0 \;\;\to\;\; A(r_0) = 1 - \sqrt{1+Q_4 (r_0)} = 0 \; .
\lab{Q4-root}
\er
The various types of horizons are as follows.
For one positive root $r_0$ of $Q_4 (r) \equiv P_4 (r) - 1$ 
\rf{Q4-def}-\rf{Q4-root}:

\begin{itemize}
\item
The single horizon $r_0 \equiv r_{\rm Schw}$ is of Schwarzschild type 
(see Fig.14 below where $(M<0, c_1 ~{\rm any},C_0>1)$, and Fig.19 below where 
$(M=0, c_1<0, c_0>1)$) for:
\be
A(r_0) = 0 \quad , \quad \pa_r A(r_0) = - \h P^\pr_4 (r_0) > 0
\; ;
\lab{Schw-horizon}
\ee
\item
The single horizon $r_0 \equiv r_{\rm dS}$ is of de Sitter type
(see Figs.1,2,8,10,20,21 below where ) for:
\be
A(r_0) = 0 \quad ,\quad
\pa_r A(r_0) = - \h P^\pr_4 (r_0) < 0 \; .
\lab{Schw-type}
\ee
\end{itemize}

For two positive roots $r^{(1)}_0 < r^{(2)}_0$ of
$Q_4 (r) \equiv P_4 (r) - 1$ \rf{Q4-def}-\rf{Q4-root}:

\begin{itemize}
\item
The two horizons are of the same type as for the Schwarzschild-de Sitter
black hole 
when:
\br
A(r^{(1,2)}_0) = 0 \quad ,\;\; \pa_r A(r^{(1)}_0) >0 \;\; ,\;\; 
\pa_r A(r^{(2)}_0)<0 \; ,
\nonu \\
{\rm i.e.} \quad P^\pr_4 (r^{(1)}_0)< 0 \quad ,\quad P^\pr_4 (r^{(2)}_0)>0 \; ,
\lab{Schw-dS-type}
\er
(see Figs.6,7 below where 
$\bigl(M>0, -4M\Bigl(\frac{c_0}{3M}\Bigr)^{3/4}\leq c_1
<-4M\Bigl(\frac{c_0-1}{3M}\Bigr)^{3/4}, c_0>1\bigr)$ ).
In the case $\bigl(M>0, c_1 <-4M\Bigl(\frac{c_0}{3M}\Bigr)^{3/4}, c_0>1\bigr)$ 
(see Fig.9 below) there are again two horizons of the same type as for the 
Schwarzschild-de Sitter black hole, however they are separated by an
intermediate forbidden region.
\item
The two horizons are of the same type as for the Reissner-Nordstr{\"o}m
black hole for:
\br
A(r^{(1,2)}_0) = 0 \quad ,\;\; \pa_r A(r^{(1)}_0)<0 \;\; ,\;\;
\pa_r A(r^{(2)}_0)>0 \; ,
\nonu \\
{\rm i.e.}\; \;
P^\pr_4 (r^{(1)}_0)>0 \quad ,\quad P^\pr_4 (r^{(2)}_0)<0 \; ,
\lab{RN-type}
\er
however, in this case the black hole exists only in a
finite-extent space region (see Fig.12 and Fig.16 where
$\bigl(M<0, c_1>4|M|\Bigl(\frac{1-c_0}{3|M|}\Bigr)^{3/4}, c_0<1\bigr)$).
\item
In the case of coallecense of the two roots of $A(r)$:
\be
A(r^{(1,2)}_0) = 0 \;\; ,\;\; r^{(1)}=r^{(2)}\; , \quad
\pa_r A(r^{(1,2)}_0) = 0 \;\;,\; {\rm i.e.}\; P^\pr_4 (r^{(1,2)}_0)=0 \; ,
\lab{extremal-type}
\ee
the horizon is of extremal black hole type (see Fig.13 and Fig.17 where
$\bigl(M<0, c_1=4|M|\Bigl(\frac{1-c_0}{3|M|}\Bigr)^{3/4}, c_0<1\bigr)$).
\end{itemize}

\section{Graphical Representations of the Class of SSS Solutions of Modified 
${\mathbf{D=4}}$ Gauss-Bonnet Gravity}
\label{graphical}

In what follows we will graphically illustrate the various possible classes 
of solutions for $A(r)$ \rf{A-sol}, focusing on de Sitter-like asymptotics of 
the latter, with or without physical singularities, including with or without 
domain walls, as well a with or without horizons
(black hole type or de Sitter cosmological type) depending on the values of
the free integration constants $(M,c_1,c_0)$.

\subsection{\textbf{Metrics Without Forbidden Regions} ($\mathbf{M>0}$)}

\subsubsection{$\mathbf{\bigl(M>0,0<c_0<1,c_1 >-4M(c_0/3M)^{3/4}\bigr)}$}


\begin{figure}[H]
\begin{center}
\includegraphics[width=8cm,keepaspectratio=true]{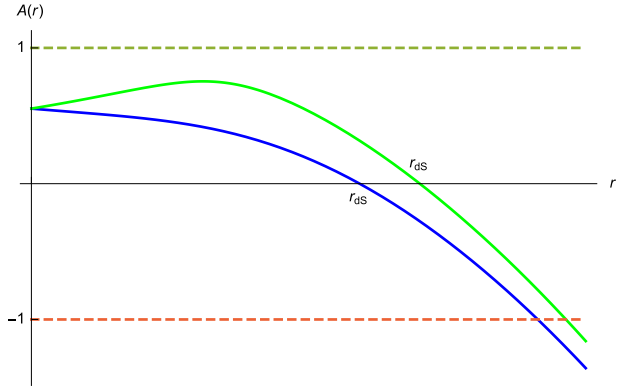}
\caption{De Sitter-like (deviation from the standard de Sitter) $A(r)$ with
a de Sitter-type horizon at $r_{\rm dS}$.
The lower curve corresponds to $c_1>0$ and the upper curve corresponds 
to $-4M(c_0/3M)^{3/4}<c_1<0$.}
\end{center}
\end{figure}

\subsubsection{$\mathbf{\bigl(M>0,0<c_0<1,c_1 =-4M(c_0/3M)^{3/4}\bigr)}$}


\begin{figure}[H]
\begin{center}
\includegraphics[width=8cm,keepaspectratio=true]{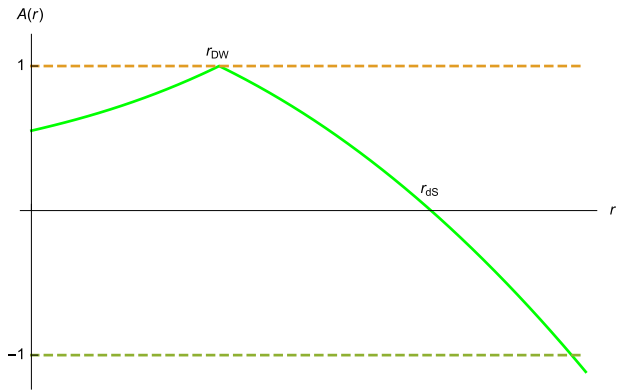}
\caption{de Sitter-like $A(r)$ with a domain wall at 
$r=r_{\rm DW}\equiv \bigl(\frac{c_0}{3M}\bigr)^{1/4}$ and a de Sitter-type
horizon at $r_{\rm dS}$. As pointed out above after Eq.\rf{brane-pressure}
the thin-shell matter of the domain wall must be {\em exotic}.}
\end{center}
\end{figure}


\subsubsection{$\mathbf{\bigl(M>0,c_0>1,c_1>0\bigr)}$}

In this case $A(r)=1-\sqrt{P_4(r)}<0$ for all $r$, so the metric is given by
\rf{K-S}, where the solution of $dr/d\xi = \sqrt{|A\bigl(r(\xi)\bigr)|}$ is
$r(\xi) \simeq \sqrt{\sqrt{c_0}-1}\,\xi$ for small $\xi$, and
$r(\xi) \simeq \exp\{M^{1/4}\xi\}$ for large $\xi$, and thus \rf{K-S}
describes monotonically expanding Kantowski-Sachs universe.

\begin{figure}[H]
\begin{center}
\includegraphics[width=8cm,keepaspectratio=true]{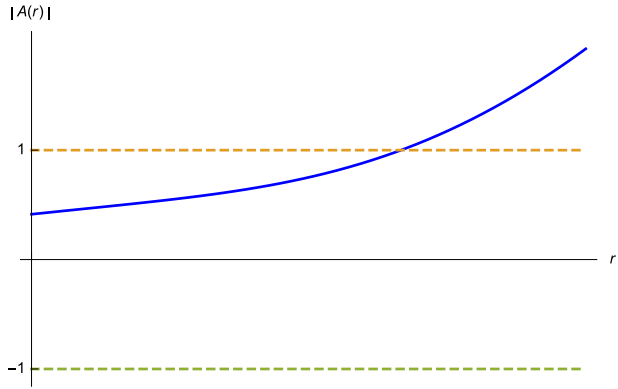}
\caption{Monotonically expanding Kantowski-Sachs universe; $r$ is timelike
coordinate.}
\end{center}
\end{figure}

\subsubsection{$\mathbf{\bigl(M>0,c_0>1,
0>c_1>-4M\bigl((c_0 -1)/3M\bigr)^{3/4}\bigr)}$}

Here again $A(r)=1-\sqrt{P_4(r)}<0$ for all $r$, but now there is a
local minimum of $|A(r)|$ at:
\be
r=r_{\rm b} \equiv \bigl(\frac{|c_1|}{4M}\bigr)^{1/3} \; ,
\lab{bounce}
\ee
as depicted on Fig.4. Therefore, the metric \rf{K-S} describes a
{\em bouncing} Kantowski-Sachs universe.

\begin{figure}[H]
\begin{center}
\includegraphics[width=8cm,keepaspectratio=true]{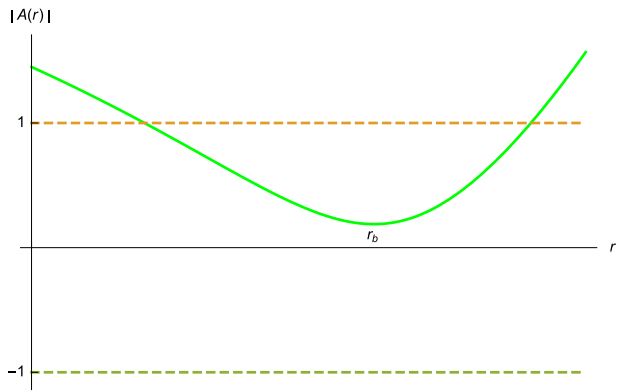}
\caption{Bouncing Kantowski-Sachs universe at $r=r_{\rm b}$; $r$ is timelike
coordinate.}
\end{center}
\end{figure}

Namely, the size squared $|A\bigl(r(\xi)\bigr)|$ of the new radial dimension $z$ 
in \rf{K-S} starts at a finite value $|A(0)|=\sqrt{\sqrt{c_0}-1}$ for $\xi=0$, 
drops down to a minimum value $|A(r_{\rm b})|$ at some finite 
cosmological time $\xi_{\rm b}$ and then it expands indefinitely for 
$\xi >\xi_{\rm b}$.

\subsubsection{$\mathbf{\bigl(M>0,c_0>1,
c_1 = -4M\bigl((c_0 -1)/3M\bigr)^{3/4}\bigr)}$}

This is a limiting case of the above bouncing Kantowski-Sachs solution, 
where now the minimum of the scale factor squared $|A(r)|$ vanishes when 
$r$ reaches $r_{\rm b}$:
\be
|A(r_{\rm b})| = 0 \quad ,\quad 
r_{\rm b} \equiv \bigl(\frac{c_0 -1}{3M}\bigr)^{1/4} \; .
\lab{BB-BC}
\ee
Accordingly we have a a very different
properies of the pertinent Kantowski-Sachs universe.

\begin{figure}[H]
\begin{center}
\includegraphics[width=8cm,keepaspectratio=true]{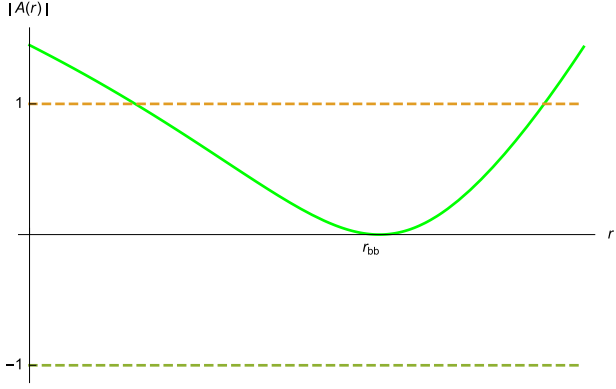}
\caption{Kantowski-Sachs universe: the left branch describing a contracting
universe with a big crunch on the interval $(0<r<r_{\rm b})$, and the right
branch describing an expanding universe with a big bang
on the interval $(r_{\rm b}<r<\infty)$; $r$ is timelike coordinate.}
\end{center}
\end{figure}

In the present case the solution $r(\xi)$ of the second
Eq.\rf{K-S} splits into two branches:

\begin{itemize}
\item
(a) Contracting Kantowski-Sachs universe with a {\em big crunch} on the
interval $0 \leq r \leq r_{\rm b}$:
\br
r(\xi) \simeq \left\{ \begin{array}{ll} {\sqrt{|A(0)|}\,\xi} \;\; & \quad {\xi
\sim 0 \; ,\; r \sim 0} \\ {r_{\rm b} - e^{-3Mr^2_{\rm b}\,\xi}} \;\; & \quad 
{\xi \to \infty \; ,\; r\sim r_{\rm b}} \end{array} \right.
\lab{r-xi-BC}
\er
Here the evolution starts at cosmological time $\xi=0$ with a non-zero scale
factor squared $|A(0)|=\sqrt{\sqrt{c_0}-1}$ (``emergent universe'')
and monotonically contracts to a 
big crunch $|A\bigl(r(\xi)\bigr)| \to |A(r_{\rm b})|= 0$ at $\xi \to \infty$.
\item
(b) Expanding Kantowski-Sachs universe with a {\em big bang} on the
interval $r_{\rm b} \leq r < \infty$:
\br
r(\xi) \simeq \left\{ \begin{array}{ll} 
{r_{\rm b} + e^{3Mr^2_{\rm b}\,\xi}} \;\; & \quad 
{\xi \to -\infty \; ,\; r\sim r_{\rm b}} \\ {e^{M^{1/4}\,\xi}} \;\; & 
\quad {\xi \to \infty\; ,\; r \to \infty} \end{array} \right.
\lab{r-xi-BB}
\er
Here evolution starts with a big bang at $\xi \to -\infty$ where the scale
factor squared $|A\bigl(r(\xi)\bigr)| \to |A(r_{\rm b})|=0$ and then
monotonically expands indefinitely for $\xi \to +\infty$.
\end{itemize}

\subsubsection{$\mathbf{\bigl(M>0,c_0>1,
-4M\bigl((c_0 -1)/3M\bigr)^{3/4}\bigr) >c_1>-4M\bigl(c_0/3M\bigr)^{3/4}\bigr)}$}



\begin{figure}[H]
\begin{center}
\includegraphics[width=8cm,keepaspectratio=true]{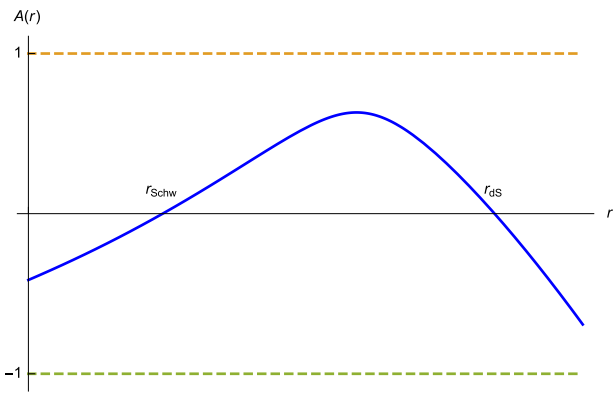}
\caption{Blackhole with two horizons -- internal Schwarzschild-type at 
$r_{\rm Schw}$ and external de Sitter-type at $r_{\rm dS}$.
Unlike the standard Schwarzschild-de Sitter
blackhole here $A(r)$ is finite at $r=0$ ($A(0)= -(\sqrt{c_0}-1)$).}
\end{center}
\end{figure}

\subsubsection{$\mathbf{\bigl(M>0,c_0>1, c_1 =-4M\bigl(c_0/3M\bigr)^{3/4}\bigr)}$}



\begin{figure}[H]
\begin{center}
\includegraphics[width=8cm,keepaspectratio=true]{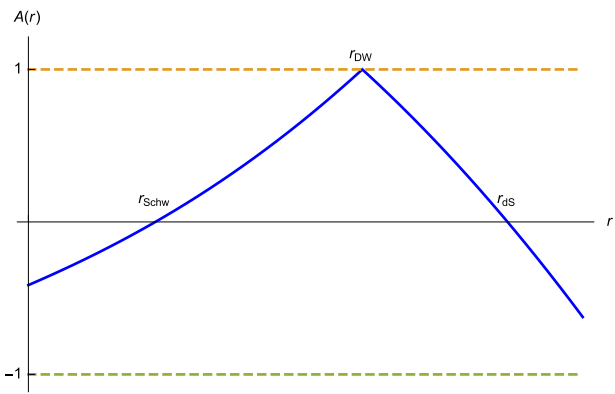}
\caption{Blackhole with two horizons -- internal Schwarzschild-type at 
$r_{\rm Schw}$ and external de Sitter-type at $r_{\rm dS}$, and with an 
additional domain wall at $r_{\rm DW}\equiv (c_0/3M)^{1/4}$ between the
two horizons. As in subsubsection 5.1.2, in particular, an additional thin-shell 
brane {\em exotic} matter source must be present to match the delta-function 
singularities in the corresponding Einstein equations at $r=r_{\rm DW}$.}
\end{center}
\end{figure}


\subsection{Metrics With One Forbidden Region ($\mathbf{M>0}$)}

\subsubsection{$\mathbf{\bigl(M>0,0<c_0<1,c_1<-4M\bigl(c_0/3M\bigr)^{3/4}\bigr)}$}

\begin{figure}[H]
\begin{center}
\includegraphics[width=8cm,keepaspectratio=true]{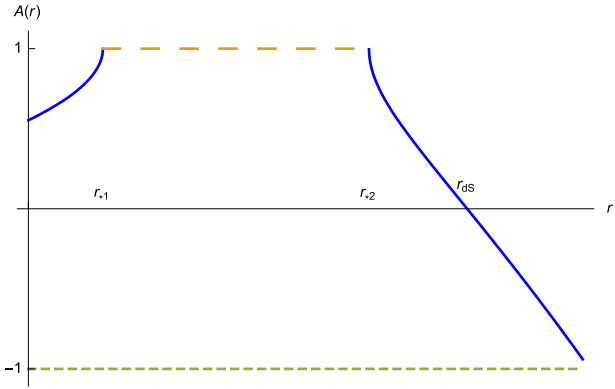}
\caption{De Sitter-like geometry with one intermediate finite-extent forbidden region
$(r_{1*}<r<r_{2*})$, and a de Sitter-type horizon at $r_{\rm dS})$.}
\end{center}
\end{figure}

\subsubsection{$\mathbf{\bigl(M>0,c_0>1,c_1<-4M\bigl(c_0/3M\bigr)^{3/4}\bigr)}$}

\begin{figure}[H]
\begin{center}
\includegraphics[width=8cm,keepaspectratio=true]{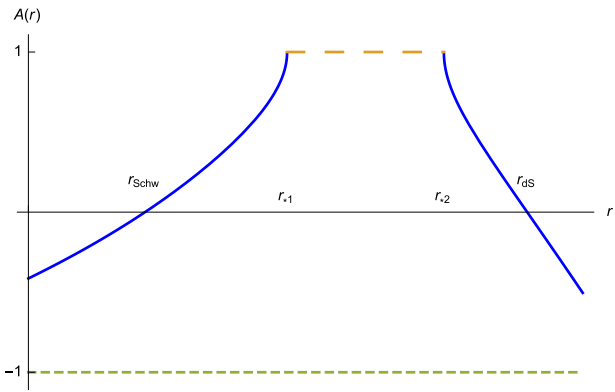}
\caption{Black Hole with two horizons -- a Schwarzschild-type at $r_{\rm Schw}$
and a de Sitter-type at $r_{\rm dS}$ separated by an intermediate finite-extent 
forbidden region $(r_{1*}<r<r_{2*})$.}
\end{center}
\end{figure}

\subsubsection{$\mathbf{\bigl(M>0,c_0<0,c_1\; {\rm any}\bigr)}$}

\begin{figure}[H]
\begin{center}
\includegraphics[width=8cm,keepaspectratio=true]{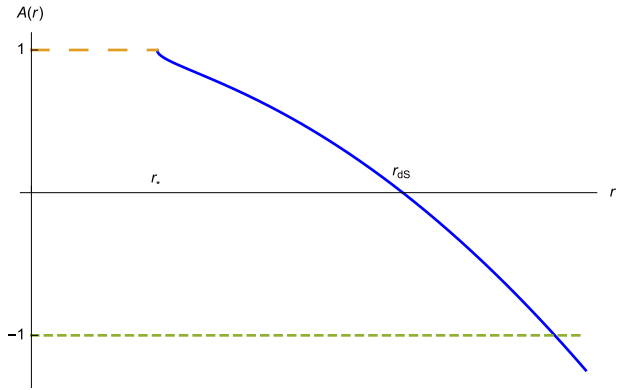}
\caption{De Sitter-like geometry with a de Sitter-type horizon at $r_{\rm dS}$,
and one internal finite-extent forbidden region $(0<r<r_{*})$.}
\end{center}
\end{figure}

\subsection{Metrics With One Forbidden Region ($\mathbf{M<0}$)}

\subsubsection{$\mathbf{\bigl(M<0,0<c_0<1,
c_1<4|M|\bigl((1-c_0)/3|M|\bigr)^{3/4}\bigr)}$}

\begin{figure}[H]
\begin{center}
\includegraphics[width=8cm,keepaspectratio=true]{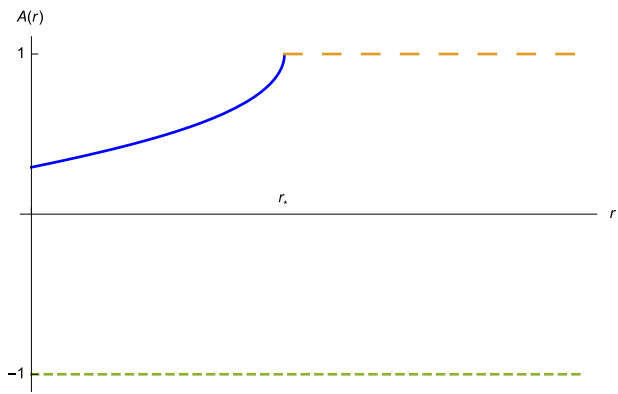}
\caption{Finite-extent internal region of regular geometry $(0<r<r_{*})$, and
one external infinite forbidden region $(r_{*}<r<\infty)$.}
\end{center}
\end{figure}

\subsubsection{$\mathbf{\bigl(M<0,0<c_0<1,
c_1>4|M|\bigl((1-c_0)/3|M|\bigr)^{3/4}\bigr)}$}

\begin{figure}[H]
\begin{center}
\includegraphics[width=8cm,keepaspectratio=true]{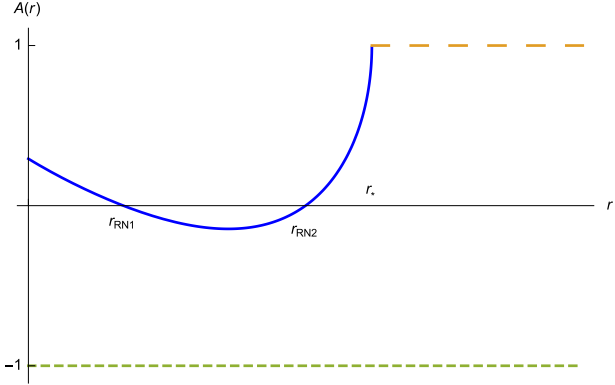}
\caption{Black hole with two horizons of Reissner-Nordstr{\"o}m-type
at $r_{\rm RN1}$ and $r_{\rm RN2}$
within in a finite extent internal space region $(0<r<r_{*})$, and an
infinite external forbidden region $(r_{*}<r<\infty)$.}
\end{center}
\end{figure}

\subsubsection{$\mathbf{\bigl(M<0,0<c_0<1,
c_1=4|M|\bigl((1-c_0)/3|M|\bigr)^{3/4}\bigr)}$}

\begin{figure}[H]
\begin{center}
\includegraphics[width=8cm,keepaspectratio=true]{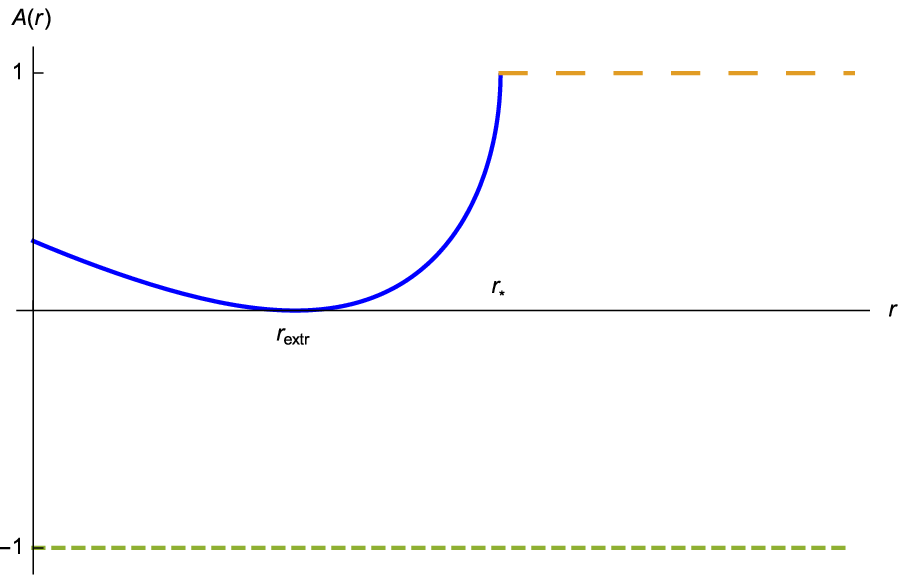}
\caption{Extremal black hole with two coallescing horizons of 
Reissner-Nordstr{\"o}m-type at $r_{\rm extr}$ within in a finite extent 
internal space region $(0<r<r_{*})$, and an infinite external forbidden 
region $(r_{*}<r<\infty)$.}
\end{center}
\end{figure}

\subsubsection{$\mathbf{\bigl(M<0,c_0>1,c_1 \; {\rm any}\bigr)}$}

\begin{figure}[H]
\begin{center}
\includegraphics[width=8cm,keepaspectratio=true]{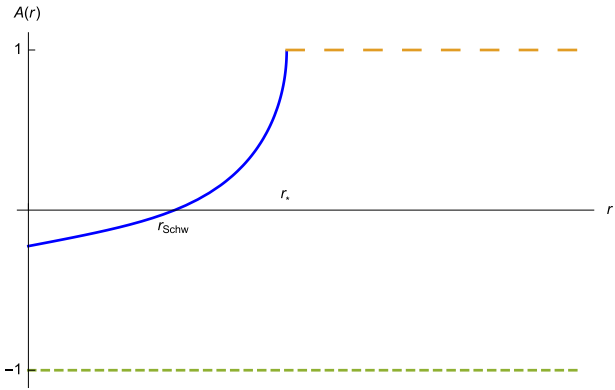}
\caption{Black hole in a finite-extent internal space region $(0<r<r_{*}$ with a
Schwarzschild-type horizon at $r_{\rm Schw}$, and with an infinite 
external forbidden region $(r_{*}<r<\infty)$.}
\end{center}
\end{figure}

\subsection{Metrics With Two Forbidden Regions ($\mathbf{M<0}$)}

\subsubsection{$\mathbf{\bigl(M<0,c_0<0,
c_1<4|M|\bigl((1+|c_0|)/3|M|\bigr)^{3/4} \bigr)}$}

\begin{figure}[H]
\begin{center}
\includegraphics[width=8cm,keepaspectratio=true]{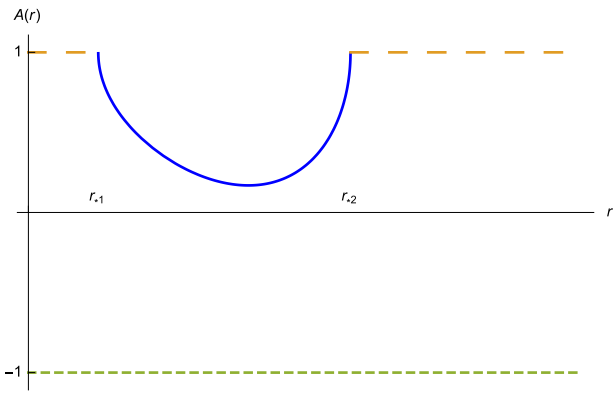}
\caption{Finite-extent intermediate space region with regular geometry
$(r_{1*}<r<r_{2*})$, and with two forbidden regions 
-- one finite-extent internal $(0<r<r_{1*}$ and one 
infinite external $(r_{2*}<r<\infty)$.}
\end{center}
\end{figure}

\subsubsection{$\mathbf{\bigl(M<0,c_0<0,
c_1>4|M|\bigl((1+|c_0|)/3|M|\bigr)^{3/4} \bigr)}$}

\begin{figure}[H]
\begin{center}
\includegraphics[width=8cm,keepaspectratio=true]{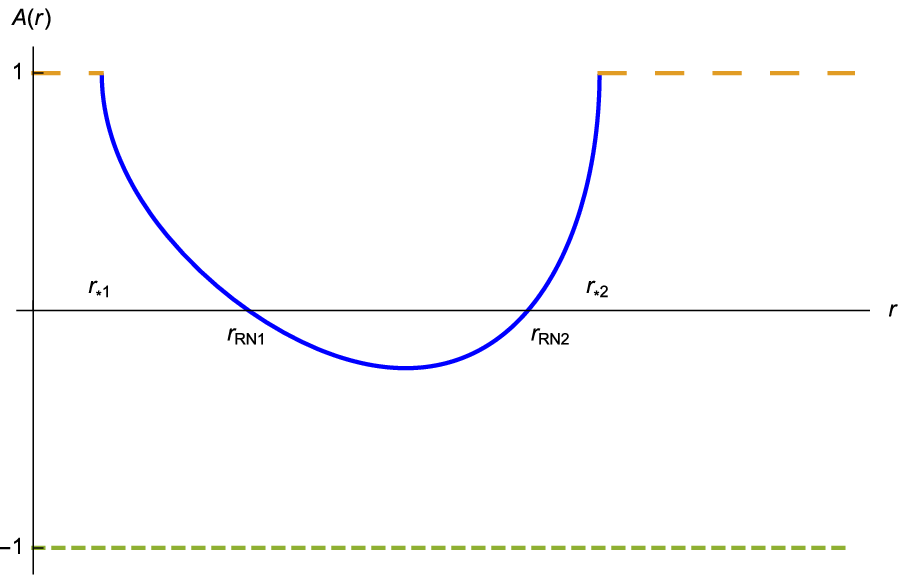}
\caption{Blackhole with two horizons of Reissner-Nordstr{\"o}m-type within a 
finite-extent intermediate space region $(r_{1*}<r<r_{2*})$, and
with two forbidden regions -- one finite-extent internal $(0<r<r_{1*}$ and one 
infinite external $(r_{2*}<r<\infty)$.}
\end{center}
\end{figure}

\subsubsection{$\mathbf{\bigl(M<0,c_0<0,
c_1=4|M|\bigl((1+|c_0|)/3|M|\bigr)^{3/4} \bigr)}$}

\begin{figure}[H]
\begin{center}
\includegraphics[width=8cm,keepaspectratio=true]{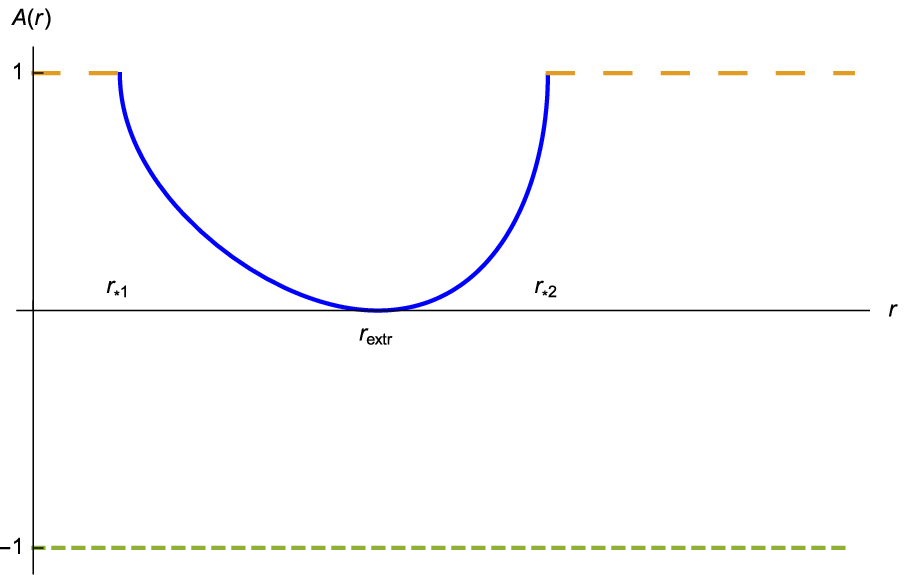}
\caption{Extremal blackhole with two coallescing horizons of 
Reissner-Nordstr{\"o}m-type within a finite-extent intermediate space region
$(r_{1*}<r<r_{2*})$, and with two forbidden regions -- one finite-extent 
internal $(0<r<r_{1*}$ and one infinite external $(r_{2*}<r<\infty)$.}
\end{center}
\end{figure}

\subsection{Metrics with $\mathbf{M=0}$}

\subsubsection{$\mathbf{\bigl(M=0,c_0>1,c_1>0\bigr)}$}

Here again $A(r)=1-\sqrt{P_4(r)}<0$ for all $r$ and the metric \rf{K-S}
describes in this case a slowly expanding Kantowski-Sachs universe: 
$|A\bigl(r(\xi)\bigr)| \sim \xi^{2/3}$ for $\xi \to \infty$.

\begin{figure}[H]
\begin{center}
\includegraphics[width=8cm,keepaspectratio=true]{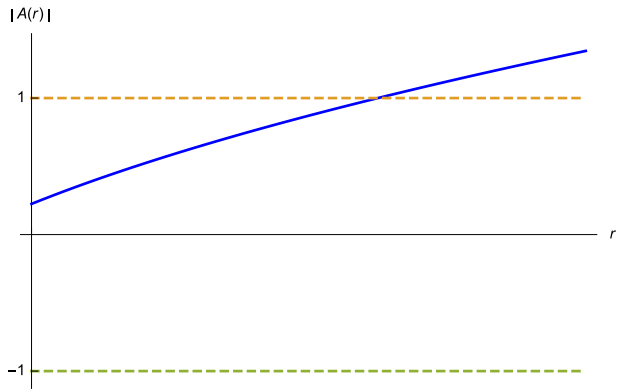}
\caption{Kantowski-Sachs slowly expanding universe.}
\end{center}
\end{figure}

\subsubsection{$\mathbf{\bigl(M=0,c_0>1,c_1<0\bigr)}$}

\begin{figure}[H]
\begin{center}
\includegraphics[width=8cm,keepaspectratio=true]{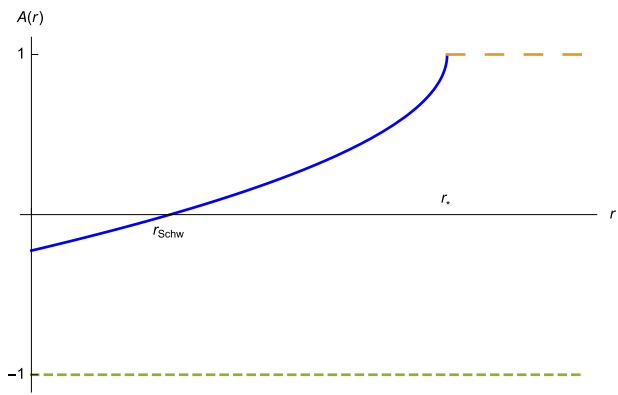}
\caption{Black hole with a Schwarzschild-type horizon within a finite-extent 
internal space region $(0<r<r_{*})$, and with an infinite external forbidden 
region $(r_{*}<r<\infty)$.}
\end{center}
\end{figure}

\subsubsection{$\mathbf{\bigl(M=0,0<c_0<1,c_1>0\bigr)}$}

\begin{figure}[H]
\begin{center}
\includegraphics[width=8cm,keepaspectratio=true]{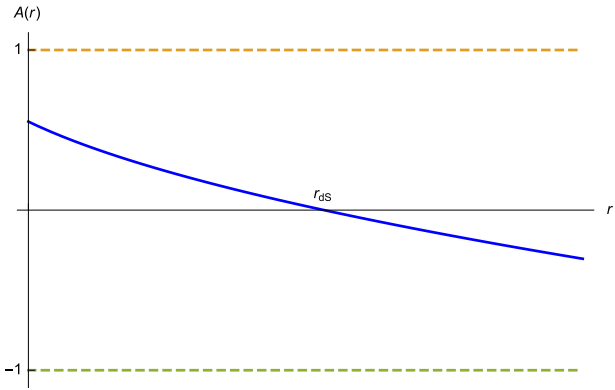}
\caption{De Sitter-like geometry with a de-Sitter type horizon at $r_{\rm dS}$.}
\end{center}
\end{figure}

\subsubsection{$\mathbf{\bigl(M=0,c_0<0,c_1>0\bigr)}$}

\begin{figure}[H]
\begin{center}
\includegraphics[width=8cm,keepaspectratio=true]{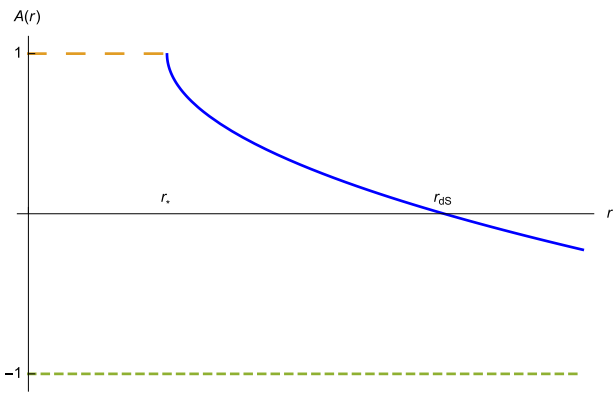}
\caption{De Sitter-like geometry in an external space region $(r_{*}<r<\infty)$
with a de Sitter type horizon at $r_{\rm dS}$,
and with one finite-extent internal forbidden region $(0<r<r_{*})$.}
\end{center}
\end{figure}

\section{Avoiding Spacetime Singularities via Domain Walls}
\label{avoid}

Let us consider two SSS solutions \rf{A-sol} with forbidden regions:

$\phantom{aaaa}$(a) The one depicted on Fig.9 -- describing 
black hole with two horizons (one Schwarzschild-type and one de Sitter-type)
separated by an intermediate finite-extent forbidden region 
$(r_{*1} < r < r_{*2})$, where $r_{*1,2}$ are two roots of of 
the 4th-order polynomial under the square root in::
\br
A_1 (r) = 1 - \sqrt{Mr^4 + c_1^{(1)}r + c_0^{(1)}} \; ,
\lab{A-2} \\
\bigl(M>0, c_0^{(1)} >1, c_1^{(1)}< -4M(c_0/3M)^{3/4}\bigr) \; .
\nonu
\er

$\phantom{aaaa}$(b) The second one graphically depicted on Fig.10 -- 
describing de Sitter-like geometry with a de Sitter-type horizon, and 
with one internal
finite-extent forbidden region $(0 < r < r_{*})$, where $r_{*}$ is a root of
the 4th-order polynomial under the square root in:
\be
A_2 (r) = 1 - \sqrt{Mr^4 + c_1^{(2)}r + c_0^{(2)}} \quad ,\;\;
\bigl(M>0, c_0^{(2)} <0, c_1^{(2)} ~{\rm any}\bigr) \; .
\lab{A-1}
\ee

Now, we can construct another SSS solution ${\widehat A}(r)$ 
{\em without any spacetime singularities} by picking a point $r={\widehat r}$ with 
${\widehat r} < r_{1*}$ ($r_{1*}$ from \rf{A-1}) and ${\widehat r} > r_{*}$
($r_{*}$ from \rf{A-2}), and glue together $A_1 (r)$ and $A_2 (r)$ at 
$r={\widehat r}$:
\br
{\widehat A}(r) =  \left\{\begin{array}{ll} A_1 (r) \;\; ,
& \quad 0<r \leq {\widehat r} \\
A_2 (r) \;\; , & \quad {\widehat r} \leq r <\infty
\end{array} \right. \;,
\lab{A-glued}
\er
so that ${\widehat A}(r)$ is continuous at $r={\widehat r}$:
\be
A_1({\widehat r}) = A_2({\widehat r}) \;\; \longrightarrow \;\;
{\widehat r} = \frac{c_0^{(1)} - c_0^{(2)}}{c_1^{(2)} - c_1^{(1)}} \; ,
\lab{A-hat-cont}
\ee
but its first derivative has a discontinuity at $r={\widehat r}$:
\be
\bigl\lb \pa_r {\widehat A} \bigr\rb_{{\widehat r}} \equiv
\pa_r A_2 ({\widehat r}) - \pa_r A_1 ({\widehat r}) = 
- \frac{c_1^{(2)} - c_1^{(1)}}{2 
\sqrt{M{\widehat r}^4 + c_1^{(2)}{\widehat r} + c_0^{(2)}}} \; ,
\lab{A-hat-deriv}
\ee
and thus the second derivative acquires delta-function contribution 
at $r={\widehat r}$:
\be
\pa_r^2 {\widehat A} = \bigl\lb \pa_r {\widehat A} \bigr\rb_{{\widehat r}}
\d(r-{\widehat r}) + \ldots  \; .
\lab{A-hat-2nd-deriv}
\ee
As in the case of \rf{A-delta} above, Eq.\rf{A-hat-2nd-deriv} imply presence of
thin-shell generated domain wall at $r={\widehat r}$ where the corresponding
brane surface tension matching the delta-function term in \rf{A-hat-2nd-deriv} is 
(cf. \rf{T-brane} and \rf{brane-pressure}):
\be
T^\th_\th = S^\th_\th \d(r-{\widehat r}) \quad ,\quad 
S^\th_\th = \bigl\lb \pa_r {\widehat A} \bigr\rb_{{\widehat r}} \; ,
\lab{DW-pressure}
\ee
with $\bigl\lb \pa_r {\widehat A}\bigr\rb_{{\widehat r}}$ as defined in 
\rf{A-hat-deriv}. Note that the latter is negative, therefore so is the
brane surface tension $S^\th_\th$, which confirms the {\em exotic} nature of
the domain wall brane matter, as already pointed out above after 
Eq.\rf{brane-pressure}.

The graphical
representation of \rf{A-glued} (Fig.22) is completely analogous to the
case of Fig.7 above 
where $\bigl(M>0, c_0>1, c_1 = - 4M (c_0/3M)^{3/4}\bigr)$.

\begin{figure}[H]
\begin{center}
\includegraphics[width=8cm,keepaspectratio=true]{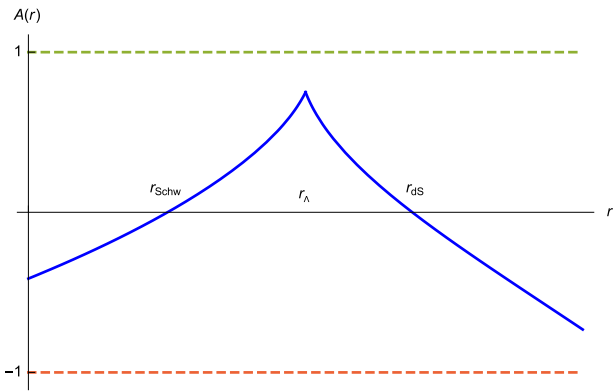}
\caption{Blackhole with two horizons -- internal Schwarzschild-type at 
$r_{\rm Schw}$ and external de Sitter-type at $r_{\rm dS}$, and with an 
additional domain wall at ${\widehat r}$ between the two horizons.
The section of the graphics for $0<r<{\widehat r}$ is as in Fig.9, whereas
the section for ${\widehat r}<r<\infty$ is as in Fig.10. Both forbidden regions in Fig.9 and Fig.10 are avoided.}
\end{center}
\end{figure}

The above described ``cut and glue together'' procedure closely resembles
the ``cut and paste'' formalism for constructing timelike thin-shell wormholes
in Ref.\ct{visser-book}, ch.15.

\section{Discussion and Outlook}
\label{conclude}

In the present paper we have studied in some detail the full class of static
spherically symmetric (SSS) solutions in the recently proposed by us new modified
Gauss-Bonnet gravity in $D=4$ based on the formalism of non-Riemannian spacetime 
volume-forms which avoids the need to couple the Gauss-Bonnet scalar to
matter fields or to employ higher powers of the latter as in ordinary $D=4$
Einstein-Gauss-Bonnet gravity models, where the latter couplings are needed
to avoid the ordinary $D=4$ Gauss-Bonnet density to become total derivative.

The dynamically triggered constancy of the Gauss-Bonnet density due to the
equations of motion resulting from the non-Riemannian spacetime volume
element by itself completely determines the solutions for the SSS metric
component function $g_{00}=-A(r)$ parametrized by three free integration constants.
Depending on the signs and values of the latter one finds SSS solutions with
deformed (anti)-de Sitter geometry, black holes of Schwarzschild-de Sitter
type, domain walls and Kantowski-Sachs universes (expanding, contracting and 
bouncing), as well as a multitude of SSS solutions exhibiting physical spacetime
singularities {\em not hidden behind horizons}, which border finite-extent
or infinitely large forbidden space regions. 

According to the cosmic sensorship principle \ct{penrose-69} the above
class of SSS solutions with naked (visible) spacetime singularities should 
be ruled out as physically acceptible solutions.
However, we showed that it is possible to avoid the singularities by
inserting appropriate domain walls and pairwise matching solutions with
singularities along the domain wall (a procedure analogous to the construction of
timelike thin-shell wormholes in Ref.\ct{visser-book}).

In various cases the field-theoretic Lagrangian actions of the corresponding 
matter sources for the above SSS gravity solutions are identified -- as a
complicated nonlinear electrodynamics with a {\em non-analytic} functional
dependence on $F^2$ (the square of the Maxwell tensor), and in a special
case --  as the $SO(3)$ nonlinear sigma model (the ``hedgehog'' scalar field 
\ct{eduardo-rabinowitz}). 

An important next task is to study SSS solutions in the more general setting
when the composite field $\chi \equiv \frac{\P(C)}{\sqrt{-g}}$ \rf{chi-def}
will not be ``frozen'' to a constant, \textsl{i.e.}, when one needs to solve
the full modified Einstein equations \rf{einstein-eqs} with $\chi = \chi(r)$.
Moreover, in the latter case we will need to consider the more general form 
of SSS metric than \rf{spherical-symm}:
\be
ds^2 = - A(r) dt^2 + B(r) dr^2 + r^2 \bigl(d\th^2 +\sin^2\th d\phi^2\bigr) 
\quad ,\;\; B(r) \neq A^{-1}(r) \; .
\lab{spherical-symm-gen}
\ee
Inserting the more general SSS ansatz \rf{spherical-symm-gen} into the
system of Eqs.\rf{einstein-eqs} and \rf{GB-const}, one gets a very
complicated coupled system of highly nonlinear ordinary differentional
equations of second order which clearly will require numerical treatment. 

\section*{Acknowledgements}
We gratefully acknowledge support of our collaboration through 
the academic exchange agreement between the Ben-Gurion University in Beer-Sheva,
Israel, and the Bulgarian Academy of Sciences. 
E.N. and E.G. have received partial support from European COST actions
MP-1405 and CA-16104, and from CA-15117 and CA-16104, respectively.
E.N. and S.P. are also thankful to Bulgarian National Science Fund for
support via research grant DN-18/1. 


\end{document}